\documentclass[reprint, aps, prl, floatfix,superscriptaddress,longbibliography]{revtex4-2}

\usepackage{graphicx}
\usepackage{amssymb,amsmath}
\usepackage{bm} 
\usepackage{dcolumn,dsfont}
\usepackage{subfigure}
\usepackage[T1]{fontenc} 
\usepackage{url}
\usepackage{mathrsfs}
\usepackage{slashed}
\usepackage{color}
\usepackage{verbatim}
\usepackage{enumitem}
\usepackage{txfonts}
\usepackage{soul,physics}
\usepackage[driverfallback=dvipdfm]{hyperref}
\hypersetup{pdfpagemode=FullScreen,colorlinks=true,breaklinks,urlcolor=blue,linkcolor=blue,citecolor=blue}
\usepackage{natbib}
\usepackage{environ}

\usepackage{comment}
\usepackage{tikz,fp}

\begin{document}
\title{Quantum String Interactions Revealed by Full Counting Statistics}
\author{Chang-Yan Wang}
\thanks{Contact author: changyanwang@cqu.edu.cn}
\affiliation{Department of Physics and Chongqing Key Laboratory for Strongly Coupled Physics, Chongqing University, Chongqing, 401331, China}
\author{Xue-Feng Zhang}
\thanks{Contact author: zhangxf@cqu.edu.cn}
\affiliation{Department of Physics and Chongqing Key Laboratory for Strongly Coupled Physics, Chongqing University, Chongqing, 401331, China}
\affiliation{Center of Quantum Materials and Devices, Chongqing University, Chongqing 401331, China}
\begin{abstract}
  How quantum strings interact is a basic question for extended objects in quantum many-body physics. Even the simplest hard-core constraint (no crossing), can generate a nontrivial effective potential, whose microscopic form is difficult to determine because the relative distance between the strings is intrinsically nonlocal. Here we show that this nonlocality is naturally captured by full counting statistics (FCS). For two hard-core quantum strings, we derive an analytic FCS expression for the emergent interaction by identifying the virtual process in which the two strings touch and hop back. Using the FCS--entanglement relation, we find the effective potential has the entanglement-controlled asymptotic form $\ln\Delta E(r)\sim -\pi^2 r^2/(12 S_\ell)$ up to subleading terms, where $S_\ell$ is the entanglement entropy between the two halves of a quantum string. We confirm the theory using high-precision numerical calculations and finite-size FCS estimates. Our results reveal FCS as a direct route to effective interactions between quantum topological line-defects, which may also be extended to higher-form charge. 
\end{abstract}
\maketitle

\emph{\color{blue!60}Introduction.--}
Interactions between particles bring in the complexity of the physical world. Meanwhile, more exotic phenomena seems to be highly related to the interplay among high dimensional extended objects. 
As a prototype, string-like objects appear across physics, from high-Tc superconductivity and frustrated magnetism \cite{eskes_quantizing_1996, zaanen_metallic_1998, zaanen_order_2000, weng_phase_1997, zhang_chiral_2013,zxf_shijie_prl2016,grusdt_microscopic_2019, chiu_string_2019, koepsell_imaging_2019, ho_imaging_2020,zxf_SciPostPhys2023, wang_interference_2024,zxf_Liu_prb2024_DQCP,zxf_Xiong_2025, wang_quantum_2025}, cosmology \cite{hindmarsh_cosmic_1995, kibble_topology_1976}, to flux strings in gauge theories \cite{athenodorou_closed_2011, kogut_hamiltonian_1975, luscher_anomalies_1980, luscher_symmetrybreaking_1981, wilson_confinement_1974}, and string theory \cite{green_superstring_2012, polchinski_string_1998}. Understanding how such extended objects interact is therefore the first step towards to comprehension of quantum many-string physics.

String interactions have been studied in several many-body settings. In fluctuating domain-wall and stripe systems, neighboring strings restrict each other's transverse motion and generate ``quantum entropic force" \cite{eskes_quantizing_1996, zaanen_metallic_1998, zaanen_order_2000}; quantum dimer models provide another setting where fluctuating strings and their constraints naturally appear \cite{orland_rokhsarkivelson_1994, jiang_string_2005, herzogarbeitman_solving_2019}. Yet even for a string confined by hard boundaries, different approaches have led to different asymptotic forms, reflecting the subtle nonlocal structure of the induced interaction \cite{mukhin_gas_2001, nishiyama_quantumfluctuationinduced_2002, orland_asymptotic_2005}. This problem has become increasingly timely with programmable quantum simulators, where Rydberg atom arrays and other platforms can realize constrained gauge-theory dynamics, observe string breaking, and probe fluctuating strings \cite{gonzalezcuadra_observation_2025,toric_string, muschik_u1_2017, surace_lattice_2020, xu_interplay_2026,ion_string}.

\begin{figure}[t]
  \centering
  \includegraphics[width = 0.98\linewidth]{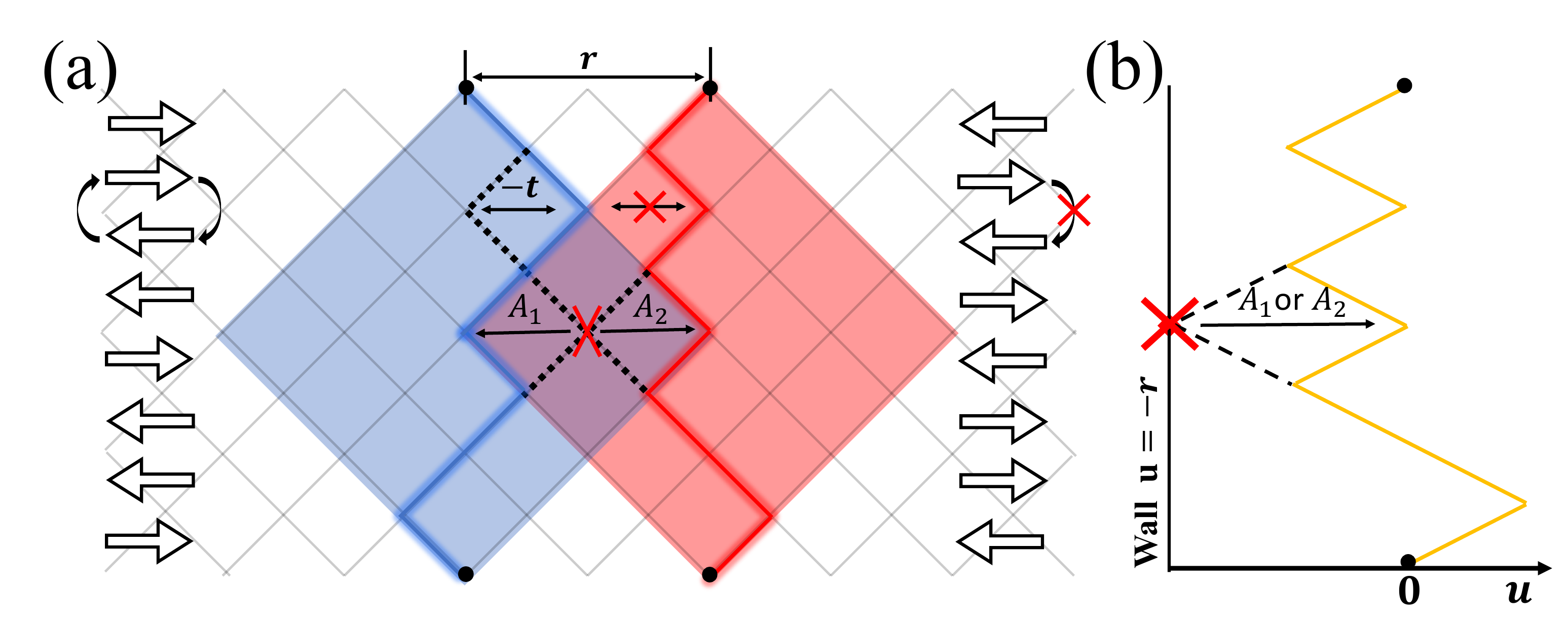}
  \caption{
  Schematic of the hard-core string interaction. (a) Local plaquette flips move two fluctuating strings separated by $r$, while the hard-core constraint forbids their crossing. The overlap of the blue and red regions marks where the strings can meet and interact. Near contact, the blocked hopping process gives the virtual wall-hopping matrix element, with return channels $A_1$ and $A_2$. Each local segment has two directions, mapped to spin up/down states. (b) In the relative-string coordinate $u$, the constraint becomes a wall at $u=-r$. The virtual process connects $u=-r$ and $u=-r+1$, so the effective boundary coordinate is centered at $u=-r+1/2$.
  }
  \label{fig:schematics}
\end{figure}

The difficulty of calculating string interactions arises from its geometric features: a classical string is a curve, and the quantum one is a worldsheet in (2+1)$d$ spacetime, which are related to the generalized symmetry \cite{generalized_symmetry_review, GLT_1form}. 
Therefore, the geometric information of the quantum strings controls their interactions. 
Full counting statistics (FCS), closely related to disorder operators \cite{kadanoff_determination_1971, fradkin_disorder_2017}, characterizes quantum numbers over extended regions and provides a natural way to retain this nonlocal information \cite{belzig_full_2001, cherng_quantum_2007, levitov_counting_2004, shelankov_charge_2003, abanov_allowed_2008, klich_measuring_2006, barratt_field_2022, bertini_nonequilibrium_2023, kitagawa_dynamics_2011,  calabrese_exact_2012, klich_manybody_2009, schonhammer_full_2007, schonhammer_full_2009, eisler_full_2013, eisler_universality_2013, ivanov_characterizing_2013, klich_note_2014, levine_full_2012, levitov_electron_1996, mcculloch_full_2023, oshima_charge_2023, klich_quantum_2009, song_bipartite_2012, song_entanglement_2011, song_general_2010, susstrunk_free_2013, wang_scaling_2021, zhao_higherform_2021, wu_universal_2021, wu_categorical_2021, liu_fermion_2023, wang_scaling_2022, tirrito_full_2023, wang_distinguishing_2024, joshi_measuring_2025, mao_probing_2025, zang_detecting_2024}. This perspective, as we will demonstrate, leads to an analytic solution for the effective potential generated by a hard-core constraint between two quantum strings.

Here we demonstrate this mechanism for two fluctuating quantum strings with a hard-core constraint. The accumulated relative displacement along the strings is encoded by a relative-string variable $u_j$. The hard-core condition then becomes the wall constraint $u_j>-r$, and the effective interaction is the constrained ground-state energy shift $\Delta E(r)=E_+(r)-E_0$. Using a Feshbach--Schur method \cite{bach_quantum_1998, feshbach_unified_1958, gustafson_mathematical_2020}, we show that $\Delta E(r)$ is governed by a virtual hopping process in which the relative string reaches the wall and hops back into the allowed region (see Fig.\ref{fig:schematics}). To make the mechanism analytically transparent, we first impose the wall only at the midpoint of the relative string. In this reduced problem, the virtual hopping matrix element becomes an exact FCS integral of the relative string. Combining this FCS result with the relation between string fluctuations and midpoint entanglement entropy, we obtain the entanglement-controlled asymptotic form
\begin{align}\label{asymptotics}
  \ln \Delta E(r)
  \sim
  -\frac{\pi^2r^2}{12S_\ell}
  \sim
  -\frac{\pi^2r^2}{2\ln L},
\end{align}
up to subleading factors, where $S_\ell$ is the entanglement entropy between the two halves of a fluctuating string.

We then restore the full wall, where the same virtual process is summed over all positions of the relative string. We show that the full-wall problem preserves the leading entanglement-controlled scale obtained from the midpoint reduction, while modifying only subleading factors and normalization. Finally, we verify the prediction using high-precision exact diagonalization (ED) and large-scale density matrix renormalization group (DMRG) calculations \cite{fishman_itensor_2021, schollwock_densitymatrix_2011, white_density_1992}. The numerical results support the asymptotic form of the effective potential in Eq.~\eqref{asymptotics}.

\emph{\color{blue!60}Model.--}
We consider two fluctuating hard-core strings separated by an integer distance $r$, with endpoints fixed on a square lattice. The string Hamiltonian is
\begin{align}
H_{\rm str}
=
-t \sum_\lozenge
\left(
\ketbra{{\color{red}<\hspace{-0.14cm}}>}{<{\color{red}\hspace{-0.14cm}>}}
+ \mathrm{h.c.}
\right),
\end{align}
where the sum runs over flippable plaquettes, as shown in Fig.~\ref{fig:schematics}(a). The hard-core constraint forbids the two strings from touching or crossing. The fluctuating regions of the two quantum strings are shaded with blue and red colors in Fig.~\ref{fig:schematics}(a), respectively.
Their overlap marks the interactive area.
This minimal model captures constrained string dynamics appearing, for example, in quantum dimer models, and can be implemented in programmable Rydberg atom arrays and trapped ion systems~\cite{gonzalezcuadra_observation_2025, surace_lattice_2020,ion_string}.

We represent the two local directions of each string segment by a spin-$1/2$ variable. Plaquette flips then become nearest-neighbor spin exchanges, giving an XY description for the unconstrained motion of each quantum string,
$
H_0
=
-t\sum_{a=1}^2\sum_{j=1}^{L-1}
(
S_j^{+,a}S_{j+1}^{-,a}
+
S_j^{-,a}S_{j+1}^{+,a}
)
$~\cite{zhang_chiral_2013}.
Fixed string endpoints impose a fixed total slope, equivalently fixed magnetization in the spin representation. For each cut $j$, we define the relative-string operator
$\hat u_j=\sum_{x=1}^{j}(S_x^{z,1}-S_x^{z,2})$, whose eigenvalue $u_j$ measures the relative displacement accumulated to the left of the cut.

In the relative-string picture, the hard-core interaction becomes a wall constraint on the allowed values of $u_j$: the relative string cannot cross the wall $u_j=-r$. Therefore, physical configurations satisfy $u_j>-r, \ j=1,\dots,L-1$. Thus, the interactive quantum strings can be described by the following Hamiltonian
$
H_+=PH_0P
$
with
$
P=\prod_{j=1}^{L-1}\Theta(\hat u_j+r)
$,
where $\Theta(x)=1$ for $x>0$ and $\Theta(x)=0$ otherwise. We call this the \emph{full-wall} problem, because the entire relative-string profile $\{u_j\}$ must remain on the allowed side of the wall $u_j=-r$. In this way, the hard-core interaction becomes an extended blocking wall in the configuration space of the relative string, as summarized in Fig.~\ref{fig:schematics}.

The ground-state energy shift defines the effective potential
$\Delta E(r)=E_+(r)-E_0$,
where $E_+(r)$ and $E_0$ are the ground-state energies of $H_+$ and $H_0$, respectively. Let $Q=1-P$ denote the projector onto the forbidden sector. Imposing the constraint removes the hopping processes that take the relative string across the wall. The leading energy correction can therefore be viewed as the self-energy generated by a virtual process from the allowed sector into the forbidden sector and back as shown in Fig.~\ref{fig:schematics}. Using the Feshbach--Schur reduction~\cite{bach_quantum_1998, feshbach_unified_1958, gustafson_mathematical_2020} with $H_0|\psi_0\rangle=E_0|\psi_0\rangle$, we obtain
\begin{align}
\Delta E(r)
=
-\frac{\mel{\psi_0}{H_{PQ}}{\psi_0}}
       {1-\mel{\psi_0}{Q}{\psi_0}},
\qquad
H_{PQ}=PH_0Q .
\end{align}
In the large-separation regime, $\mel{\psi_0}{Q}{\psi_0}$ is small, so the effective interaction is controlled by the virtual hopping matrix element $\mel{\psi_0}{H_{PQ}}{\psi_0}$. As we will see, this matrix element is resolved by the relative-string variable at the cut where the virtual hopping occurs, and its nonlocal structure is naturally captured by the FCS of the relative string.

\emph{ \color{blue!60}Mid-wall approximation.--}
The full-wall constraint involves all cuts $j=1,\dots,L-1$. To obtain a transparent analytic description, we first impose the wall only at the midpoint cut $\ell=L/2$,
\begin{align}
P_{\rm mid}=\Theta(\hat u_\ell+r),
\qquad
Q_{\rm mid}=1-P_{\rm mid}.
\end{align}
We call this the \emph{mid-wall} problem: the relative string is blocked only at $u_\ell$, the cut farthest from the fixed endpoints and therefore the one with the strongest fluctuations.

In this mid-wall problem, the virtual process is local in the relative-string coordinate. When the relative string reaches the wall at $u_\ell=-r$, a plaquette flip can return it to the allowed side. There are two such return channels, $A_1$ and $A_2$ in Fig.~\ref{fig:schematics}, corresponding to flipping the local segment of either string. Both increase $u_\ell$ by one, so the mid-wall $P$-$Q$ coupling is
\begin{align}\label{fp_eq}
H_{PQ}^{\rm mid}
=
-t(A_1+A_2)\delta_{\hat u_\ell,-r}.
\end{align}
This wall-hopping operator describes the virtual process in which the relative string reaches the wall and hops back into the allowed sector.

To evaluate its matrix element, we use the Jordan--Wigner representation. For open boundary conditions, the Jordan--Wigner strings cancel between nearest neighbors, and the unconstrained Hamiltonian becomes
$
H_0=-t\sum_{a=1}^2\sum_{j=1}^{L-1}
(c_{j,a}^\dagger c_{j+1,a}+c_{j+1,a}^\dagger c_{j,a})
$,
and $\hat u_j=\hat N_j^{(1)}-\hat N_j^{(2)}$, with $\hat N_j^{(a)}=\sum_{x=1}^{j}\hat n_{x,a}$. In this representation,
\begin{align}
A_1=c_{\ell,1}^\dagger c_{\ell+1,1},
\qquad
A_2=c_{\ell+1,2}^\dagger c_{\ell,2}.
\end{align}
The one-string ground state is the half-filled Slater determinant $|{\rm GS}\rangle$, so the unconstrained two-string ground state factorizes as $|\psi_0\rangle=|{\rm GS}\rangle_1\otimes |{\rm GS}\rangle_2$. We denote expectations in $|\psi_0\rangle$ by $\langle\cdots\rangle_0$ and those in $|{\rm GS}\rangle$ by $\langle\cdots\rangle$.

The factorized ground state reduces this matrix element to one-string
quantities. Let
$p_n=\langle\delta_{\hat N_\ell,n}\rangle$ and
$f_n=\langle c_\ell^\dagger c_{\ell+1}\delta_{\hat N_\ell,n}\rangle$ in the
one-string ground state. Since
$\delta_{\hat u_\ell,-r}=\sum_n
\delta_{\hat N_\ell^{(1)},n}\delta_{\hat N_\ell^{(2)},n+r}$, the $A_1$
channel gives $\sum_n f_np_{n+r}$.
The $A_2$ channel can be written in terms of the same one-string quantity after shifting the number label. Since the one-string ground-state wave function can be chosen real and $c_\ell^\dagger c_{\ell+1}$ raises $\hat N_\ell$ by one, this channel contributes $\sum_n p_nf_{n+r-1}$. Therefore,
\begin{align}\label{mid_hpq}
\langle H_{PQ}^{\rm mid}\rangle_0
=
-t\sum_{n\in\mathbb Z}
\left[
f_np_{n+r}+p_nf_{n+r-1}
\right],
\end{align}
which reduces the mid-wall contribution to number-resolved hopping matrix elements of a single
string at the midpoint cut.

\emph{\color{blue!60}FCS and effective potential.--}
The mid-wall matrix element admits a compact representation in terms of the FCS of the relative string. This can be seen by noting that the coefficients $p_n$ and $f_n$ are the Fourier coefficients of one-string FCS generating function $\chi(\lambda)=\langle e^{i\lambda\hat N_\ell}\rangle$ and the hopping-inserted generating function $\Gamma(\lambda)=\langle c_\ell^\dagger c_{\ell+1}e^{i\lambda\hat N_\ell}\rangle$ respectively, where $\lambda \in (-\pi,\pi]$ \cite{klich_quantum_2009, calabrese_exact_2012}. Writing $\Gamma(\lambda)=\chi(\lambda)m(\lambda)$ and substituting the Fourier representations into Eq.~\eqref{mid_hpq}, we obtain
\begin{align}\label{fcs_integral}
\langle H_{PQ}^{\rm mid}\rangle_0
=
-t\int_{-\pi}^{\pi}\frac{d\lambda}{2\pi}
e^{-ir\lambda}\mathcal F(\lambda)A(\lambda),
\end{align}
where $\mathcal F(\lambda)=\chi(\lambda)\chi(-\lambda)=\langle e^{i\lambda\hat u_\ell}\rangle_0$ is the FCS generating function of the midpoint relative-string operator $\hat u_\ell=\hat N_\ell^{(1)}-\hat N_\ell^{(2)}$, and $A(\lambda)=m(-\lambda)+e^{i\lambda}m(\lambda)$ is the hopping insertion. For the string considered here, These quantities are actually determined by the correlation matrix $C_{xy}=\langle c_x^\dagger c_y\rangle$. Explicitly, we have $\chi(\lambda)=\det[I-C+CD_\ell(\lambda)]$, where $D_\ell(\lambda)=I+(e^{i\lambda}-1)P_\ell$ and $P_\ell$ projects onto sites $1,\dots,\ell$ \cite{abanov_allowed_2008, klich_measuring_2006, sm}. The corresponding hopping-inserted determinant formula gives $m(\lambda)=[(I-C+CD_\ell)^{-1}CD_\ell]_{\ell,\ell+1}$ (details can be found in the supplementary materials \cite{sm}).

The leading asymptotic structure follows from the small-$\lambda$ behavior of Eq.~\eqref{fcs_integral}. Let $\nu_a$ be the eigenvalues of the restricted correlation matrix $C_\ell$. These eigenvalues determine both the one-string FCS, $\chi(\lambda)=\prod_a[1+(e^{i\lambda}-1)\nu_a]$ \cite{calabrese_exact_2012, klich_manybody_2009, schonhammer_full_2007, schonhammer_full_2009}, and the bipartite entanglement entropy across the midpoint, $S_\ell=-\sum_a[\nu_a\ln\nu_a+(1-\nu_a)\ln(1-\nu_a)]$ \cite{vidal_entanglement_2003, calabrese_exact_2012, jin_quantum_2004, peschel_calculation_2003, peschel_reduced_2009}. Expanding the FCS near $\lambda=0$ gives $\ln\chi(\lambda)=i\lambda\langle\hat N_\ell\rangle-\lambda^2\langle\hat N_\ell^2\rangle_c/2+O(\lambda^3)$. The odd terms cancel in $\mathcal F(\lambda)=\chi(\lambda)\chi(-\lambda)$, so
\begin{align}\label{gaussian_fcs}
\mathcal F(\lambda)
=
\exp\left[
-\frac{3S_\ell}{\pi^2}\lambda^2
+O(\lambda^4)
\right],
\end{align}
where we used the single-string relation of entanglement entropy and connected correlator
$S_\ell\simeq(\pi^2/3)\langle\hat N_\ell^2\rangle_c=(\pi^2/6)G_{\ell\ell}$, with $G_{\ell\ell}=\langle\hat u_\ell^2\rangle_c=2\langle\hat N_\ell^2\rangle_c$ \cite{klich_manybody_2009, calabrese_exact_2012}. Thus the same entanglement scale that controls the midpoint bipartite entropy also controls the Gaussian FCS factor of the relative string.

For the real single-string ground state, we have $m(-\lambda)=m(\lambda)^*$, and hence $A(\lambda)=e^{i\lambda/2}B(\lambda)$ with $B(\lambda)=2\operatorname{Re}[e^{i\lambda/2}m(\lambda)]$. Expanding the determinant expression gives $m(\lambda)=C_{\ell,\ell+1}-i\lambda(CP_\ell C)_{\ell,\ell+1}+O(\lambda^2)$ \cite{sm}. Therefore the linear correction to $e^{i\lambda/2}m(\lambda)$ is purely imaginary, and $B(\lambda)=2C_{\ell,\ell+1}+O(\lambda^2)$. Equation~\eqref{fcs_integral} becomes
\begin{align}
\langle H_{PQ}^{\rm mid}\rangle_0
=
-t\int_{-\pi}^{\pi}\frac{d\lambda}{2\pi}
e^{-i(r-1/2)\lambda}\mathcal F(\lambda)B(\lambda).
\end{align}
The hopping insertion therefore shifts the Fourier phase by one half. Keeping the Gaussian FCS factor and the leading value $B(0)=2C_{\ell,\ell+1}$ gives
\begin{align}
\langle H_{PQ}^{\rm mid}\rangle_0
\simeq
-\frac{2tC_{\ell,\ell+1}}{\sqrt{12S_\ell/\pi}}
\exp\left[
-\frac{\pi^2(r-1/2)^2}{12S_\ell}
\right],
\end{align}
up to subleading factors. The shift $r\to r-1/2$ has a simple interpretation: the virtual hopping process connects the forbidden sector $u_\ell=-r$ and the nearest allowed sector $u_\ell=-r+1$, so the matrix element is centered at the bond midpoint $u_\ell=-r+1/2$ as illustrated in Fig.\ref{fig:schematics}(b).

For the string considered here, we have $S_\ell=(1/6)\ln L+O(1)$ \cite{vidal_entanglement_2003, calabrese_entanglement_2004}, equivalently $G_{\ell\ell}=\pi^{-2}\ln L+O(1)$, a direct derivation of which is given in the End Matter. Combining these relations, we obtain the mid-wall effective potential
\begin{align}\label{midasymptotics}
\ln \Delta E_{\rm mid}(r)
\sim
-\frac{\pi^2 r^2}{12S_\ell}
\sim
-\frac{\pi^2r^2}{2\ln L},
\end{align}
up to subleading terms.

This result has a simple interpretation. If the hopping insertion is approximated by $A(\lambda)\simeq A(0)$, Eq.~\eqref{fcs_integral} reduces to the inverse Fourier transform of $\mathcal F(\lambda)=\langle e^{i\lambda\hat u_\ell}\rangle_0$, which measures the probability of the relative string touching the wall. The hopping insertion refines this picture by shifting the relevant coordinate from the wall site $u_\ell=-r$ to the bond center $u_\ell=-r+1/2$, because the virtual process connects the forbidden sector $u_\ell=-r$ and the nearest allowed sector $u_\ell=-r+1$.

This asymptotic form of the effective potential reveals the entanglement-controlled nature of the hard-core string repulsion. It is not obtained from a local interaction energy density, but from the FCS of the nonlocal relative string operator $\hat u_\ell$. The hard-core constraint is local in relative string configuration space, yet the operator that reaches the wall is an accumulated displacement over an extended segment. These nonlocal quantum string fluctuations are tied to the entangled nature of the fluctuating string and are quantified, at leading scale, by the entanglement entropy between its two halves. In this sense, the hard-core repulsion is dictated by the entanglement-controlled wandering of the string, leading directly to the scale in Eq.~\eqref{midasymptotics}.

\emph{\color{blue!60}Full-wall problem.--}
We now return to the physical full-wall constraint. The mid-wall calculation isolates the virtual hopping process at one cut. In the full-wall problem, the same process can occur at any cut $y$, but the rest of the relative string must remain on the allowed side of the wall.

For a hop across bond $y$, the operator $T_y^{(+)}=c_{y,1}^\dagger c_{y+1,1}+c_{y+1,2}^\dagger c_{y,2}$ raises $u_y$ by one. Since this hop changes only the cut variable $u_y$, it connects the forbidden side to the allowed side only when the initial value is $u_y=-r$, while all other cuts already satisfy $u_j>-r$. Therefore, the full-wall virtual hopping operator is
\begin{align}
H_{PQ}
=
-t\sum_{y=1}^{L-1}
T_y^{(+)}
\delta_{\hat u_y,-r}
\prod_{j\ne y}\Theta(\hat u_j+r).
\end{align}
This is the direct extension of the mid-wall operator. The delta function selects the cut where the virtual hopping process crosses the wall, and the product of step functions enforces the full-wall constraint on all other cuts.

The corresponding matrix element can be written using the joint FCS of the relative string. Define $\mathcal F(\boldsymbol\lambda)=\langle e^{i\sum_j\lambda_j\hat u_j}\rangle_0=\chi_L(\boldsymbol\lambda)\chi_L(-\boldsymbol\lambda)$. For one string, $\sum_j\lambda_j\hat N_j=\sum_x\alpha_x\hat n_x$ with $\alpha_x=\sum_{j=x}^{L-1}\lambda_j$, so $\chi_L(\boldsymbol\lambda)=\det[I-C+CD(\boldsymbol\lambda)]$, where $D(\boldsymbol\lambda)=\operatorname{diag}(e^{i\alpha_1},\dots,e^{i\alpha_{L-1}}, 1)$. The hopping insertion at cut $y$ is obtained from $R=(I-C+CD)^{-1}CD$ as $A_y(\boldsymbol\lambda)=R(\boldsymbol\lambda)_{y,y+1}+R(-\boldsymbol\lambda)_{y+1,y}$. Thus
\begin{align}\label{fullwall}
\langle H_{PQ}\rangle_0
=
-t\sum_{y=1}^{L-1}
\sum_{\mathbf m\in\mathcal A_r^{(y)}}
\int\frac{d^{L-1}\lambda}{(2\pi)^{L-1}}
e^{-i\boldsymbol\lambda\cdot\mathbf m}
\mathcal F(\boldsymbol\lambda)A_y(\boldsymbol\lambda),
\end{align}
where $\mathcal A_r^{(y)}$ denotes relative-string configurations with $m_y=-r$ and $m_j>-r$ for $j\ne y$ \cite{sm}.

Near $\boldsymbol\lambda=0$, the joint FCS has the Gaussian form
$
\mathcal F(\boldsymbol\lambda)
=
\exp\left[
-\frac12\boldsymbol\lambda^TG\boldsymbol\lambda
+O(\lambda^4)
\right],
\quad
G_{jk}=\langle \hat u_j\hat u_k\rangle_c .
$
The insertion is smooth and satisfies $A_y({\bf 0})=2C_{y,y+1}$. Keeping its leading phase gives $A_y(\boldsymbol\lambda)\simeq2C_{y,y+1}e^{-i\lambda_y/2}$, so only the crossed cut is shifted from $u_y=-r$ to the bond center $u_y=-r+1/2$. The remaining cuts are not shifted; they only enforce the full-wall constraint.

For a given cut $y$, the leading FCS scale is controlled by the diagonal correlator $G_{yy}$. Up to subleading factors from the constraint on the remaining cuts, this gives a contribution of the form $\exp[-(r-1/2)^2/(2G_{yy})]$. The dominant contribution therefore comes from the cut with the largest fluctuation, $G_{\max}=\max_yG_{yy}$, or equivalently from the cut with the largest single-string entanglement entropy $S_{\max}=\max_y S_y$, where $S_y$ denotes the entanglement entropy across cut $y$.
Using the leading relation $S_y\simeq(\pi^2/6)G_{yy}$, we obtain \cite{sm}
\begin{align}
\Delta E(r)
\sim
\exp\left[
-\frac{\pi^2(r-1/2)^2}{12S_{\max}}
\right]
\times \text{subleading factors}.
\end{align}
Similar to the mid-wall problem, we have $G_{\max}=\pi^{-2}\ln L+O(1)$ and $S_{\max}=(1/6)\ln L+O(1)$ near the midpoint. So the full-wall potential has the same leading scale as the mid-wall result, and we thus obtain the entanglement-controlled asymptotics of hard-core quantum string interactions Eq.~\eqref{asymptotics}.

Therefore, the hard-core interaction between two fluctuating strings is controlled by the FCS of the nonlocal relative string. The mid-wall problem exposes the local wall-hopping mechanism, while the full-wall problem sums the same virtual process over all cuts, with the rest of the string constrained to remain on the allowed side. The leading scale is still dictated by the largest entanglement-controlled string wandering, while the full-wall constraint modifies only subleading factors and normalization.

\emph{\color{blue!60}Numerical results.--}
We compute the effective potential $\Delta E(r)$ and compare it with the FCS theory. For small systems, we use high-precision ED directly in the constrained Hilbert space. For larger systems, we use DMRG for the projected Hamiltonian $H_+=PH_0P$.
Implementation details are given in the Supplemental Material \cite{sm}.

\begin{figure}[t]
  \centering
  \includegraphics[width=0.94\linewidth]{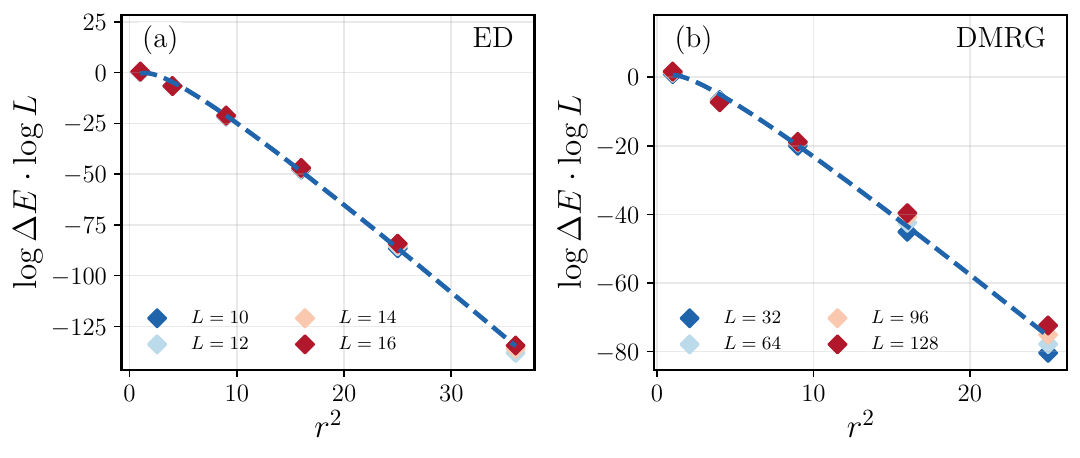}
  \caption{
  Scaling fit of $\ln\Delta E\cdot\ln L$ using $\ln\Delta E\cdot\ln L=-a(r-1/2)^2+b\ln r+c$. The dashed line is the fit. Panel (a) shows ED data for $L=10,12,14,16$, giving $a=4.974\pm0.068$. Panel (b) shows DMRG data for $L=32,64,96,128$, giving $a=4.107\pm0.156$.
  }
  \label{fig:numerical_scaling_fit}
\end{figure}

We first examine the scaling behavior of the effective potential. The FCS theory predicts the dominant asymptotics are controlled by the midpoint entanglement entropy $S_\ell\simeq(1/6)\ln L+O(1)$, together with the half-link refinement $r\to r-1/2$. We therefore fit the scaled data to
\begin{align}
\ln\Delta E \cdot \ln L
=
-a(r-1/2)^2+b\ln r+c .
\end{align}
The leading prediction is $a=\pi^2/2$, while $b\ln r+c$ captures prefactors and normalization corrections. As shown in Fig.~\ref{fig:numerical_scaling_fit}, this form organizes both the ED and DMRG data. The ED fit gives $a=4.974\pm0.068$, close to $\pi^2/2\simeq4.935$, while the DMRG fit gives $a=4.107\pm0.156$. We attribute this drift to finite-size and finite-accuracy effects in the exponentially small energy differences.

\begin{figure}[t]
  \centering
  \includegraphics[width=0.96\linewidth]{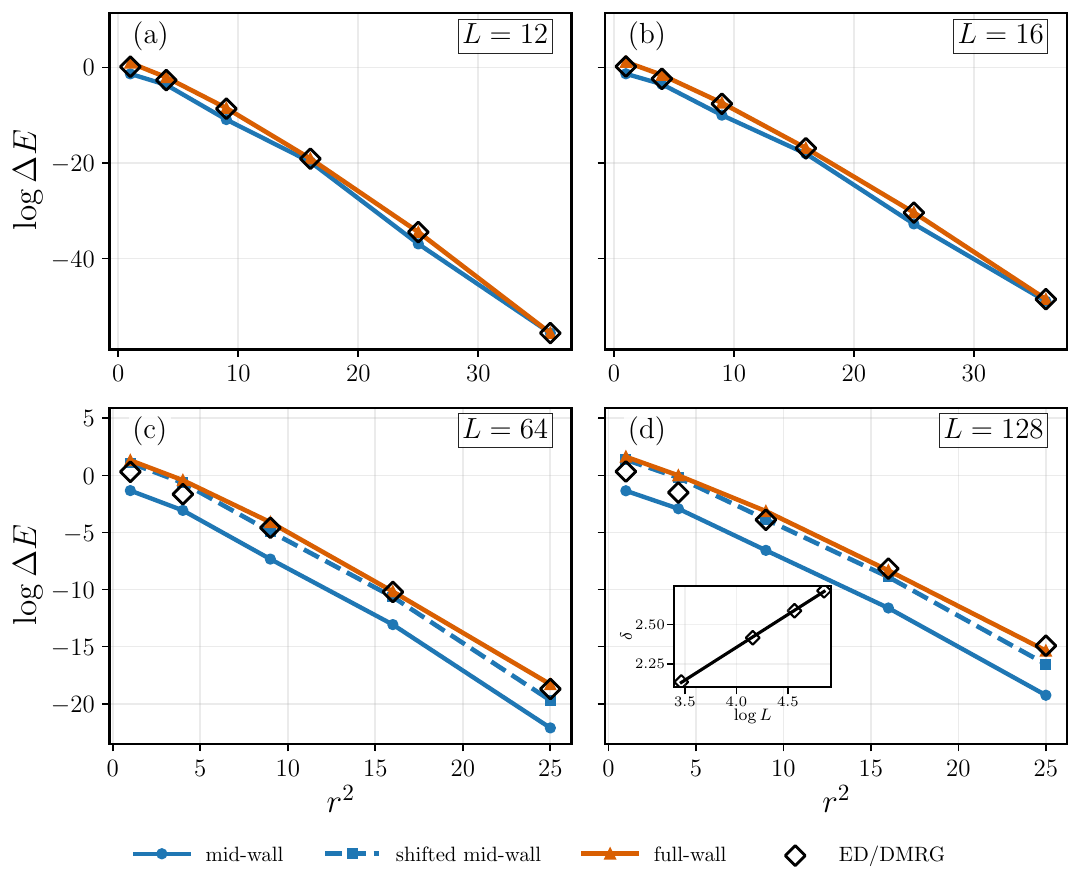}
  \caption{
  Comparison of (a,b) ED and (c,d) DMRG energies with FCS estimates. Solid blue curves are the direct finite-size mid-wall FCS integrals. Dashed blue curves for $L=64,128$ include a fitted vertical shift. Orange curves are the hybrid windowed full-wall estimates, with exact one-cut FCS factors in the central window and Gaussian conditional factors for the remaining wall constraint. The inset shows the fitted shift $\delta(L)\simeq0.417\ln L+0.687$ for $L=32,64,96,128$.
  }
  \label{fig:numerical_integral_comparison}
\end{figure}

We next compare the numerical energies with finite-size FCS estimates, as shown in Fig.~\ref{fig:numerical_integral_comparison}. The mid-wall curve is obtained from Eq.~\eqref{fcs_integral} using the exact finite-$L$ determinant expressions for $\mathcal F(\lambda)$ and $A(\lambda)$, without using the small-$\lambda$ Gaussian approximation. It agrees well with the ED data at small sizes. For larger systems, the remaining mismatch is close to a vertical shift on the logarithmic scale, indicating a slowly varying normalization correction rather than a change in the separation-dependent exponent.

To estimate the full-wall FCS formula Eq.~\eqref{fullwall}, we use the hybrid windowed FCS method shown in Fig.~\ref{fig:numerical_integral_comparison}. The one-cut FCS integral is evaluated exactly near the central cut, where $G_{yy}$ is maximal, and replaced by its Gaussian form away from this window. This keeps the dominant finite-size FCS structure while avoiding the full high-dimensional joint-FCS integral. The resulting full-wall estimate agrees well with both ED and DMRG energies on the logarithmic scale.

These numerical results support the FCS mechanism for the repulsive interaction between hard-core quantum strings. The effective potential follows the entanglement-controlled scaling variable $(r-1/2)^2/S_\ell$, the finite-size mid-wall integral reproduces the hopping-inserted FCS structure, and the windowed full-wall estimate incorporates the additional constraint on all cuts. The leading scale is therefore given by Eq.~\eqref{asymptotics}, with the half-link shift and smooth normalization corrections appearing beyond the leading exponential form.

\emph{\color{blue!60}Conclusion and Discussion.--}
We have derived a microscopic effective potential generated solely by the hard-core constraint between two quantum strings. The central result is that this repulsion is not governed by a local energy density, but by the FCS of a nonlocal relative string. Through the FCS--entanglement relation, the same collective fluctuations are measured by the entanglement entropy between two halves of a fluctuating string. Thus the hard-core constraint gives rise to an entanglement-controlled repulsion.

The resulting potential Eq.~\eqref{asymptotics} is unusual: its exponent is set by the entanglement scale of string wandering, rather than by a microscopic length. The midpoint calculation exposes the local mechanism as a virtual wall-hopping process, while the full-wall construction shows that imposing the constraint on the entire relative string preserves the same leading scale and modifies only subleading factors and normalization.

Our numerical results support this picture, showing the predicted dependence on $(r-1/2)^2/S_\ell$ and the finite-size FCS structure. More broadly, this work identifies FCS as a direct route to effective interactions between nonlocal fluctuating objects, and suggests that entanglement-controlled repulsion may also appear for quantum domain walls, stripes in high-Tc superconductivity and even classical membranes in biophysics \cite{membranes}.

{\color{blue}\it{Acknowledgement.-}} We would like to thank Tao Shi and Tie-Yan Si for many helpful discussions. X.-F. Z. acknowledges funding from the National Science Foundation of China under Grants  No.12274046 and No.12547101, and Xiaomi Foundation / Xiaomi Young Talents Program.

\bibliography{ref.bib}

\begin{widetext}

\begin{center}
{\bf End Matter}
\end{center}
\emph{\color{blue!60}Appendix.--}
Here we derive the logarithmic scale used in the main text. For one string in the free-fermion representation, let $A=\{1,\dots,\ell\}$ and $\hat N_\ell=\sum_{x\in A}\hat n_x$. The connected number fluctuation is
\begin{align}
\langle \hat N_\ell^2\rangle_c
=
\mathrm{Tr}\, C_A(1-C_A)
=
\sum_{x\le\ell}C_{xx}
-
\sum_{x,y\le\ell}|C_{xy}|^2 ,
\end{align}
where $C_A$ is the restriction of the ground-state correlation matrix $C$ to $A$. Since the full correlation matrix is a projector, $C^2=C$, this can also be written as the cross-cut sum
\begin{align}
\langle \hat N_\ell^2\rangle_c
=
\sum_{x=1}^{\ell}\sum_{y=\ell+1}^{L}|C_{xy}|^2 .
\end{align}
For the single string with two ends fixed, the exact single-particle eigenmodes give
\begin{align}
C_{xy}
=
\frac{2}{L+1}
\sum_{m=1}^{L/2}
\sin\frac{\pi mx}{L+1}
\sin\frac{\pi my}{L+1}.
\end{align}
Near the midpoint cut, the leading singular part is the bulk kernel
$C_{xy}\sim \sin[\pi(x-y)/2]/[\pi(x-y)]$, while the image contribution only changes the $O(1)$ term. Therefore the cross-cut sum reduces, at leading order, to a harmonic sum. Writing $m=y-x$, the half-filling factor $\sin^2(\pi m/2)$ selects odd separations, and
\begin{align}
\sum_{m=a}^{\infty}\frac{\sin^2(\pi m/2)}{m^2}
=
\frac{1}{2a}+O(a^{-2}).
\end{align}
Summing over the distance $a$ from the cut gives
\begin{align}
\langle \hat N_\ell^2\rangle_c
=
\frac{1}{2\pi^2}\ln L+O(1).
\end{align}
For the relative string, the two strings are independent in the unconstrained ground state, so
\begin{align}
G_{\ell\ell}
=
\langle \hat u_\ell^2\rangle_c
=
2\langle \hat N_\ell^2\rangle_c
=
\frac{1}{\pi^2}\ln L+O(1).
\end{align}

\begin{figure}[t]
  \centering
  \includegraphics[width=0.5\linewidth]{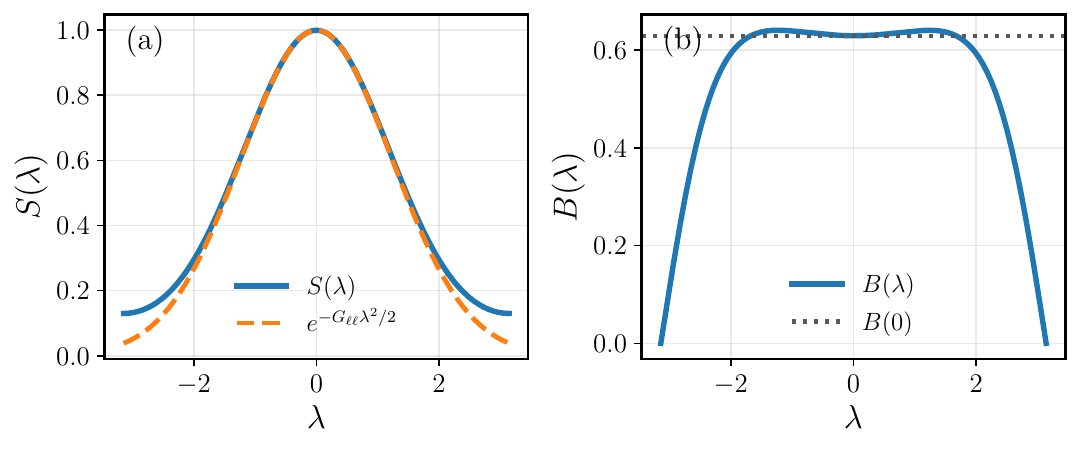}
  \caption{
  Small-$\lambda$ structure of the mid-wall FCS. (a) Exact $\mathcal F(\lambda)$ and the Gaussian approximation $e^{-G_{\ell\ell}\lambda^2/2}$. (b) Exact $B(\lambda)$ and the dotted line marks $B(0)=2C_{\ell,\ell+1}$.
  }
  \label{fig:small_lambda_structure}
\end{figure}

The relation to entanglement follows from the same restricted correlation matrix. If $\nu_a$ are the eigenvalues of $C_A$, the entanglement entropy between the two pieces of one string is
\begin{align}
S_\ell
=
-\sum_a
\left[
\nu_a\ln\nu_a
+
(1-\nu_a)\ln(1-\nu_a)
\right],
\end{align}
whereas the number fluctuation is
\begin{align}
\langle \hat N_\ell^2\rangle_c
=
\sum_a \nu_a(1-\nu_a).
\end{align}
For a one-dimensional free-fermion ground state, the second cumulant is the only cumulant with the leading logarithmic growth; higher even cumulants contribute only $O(1)$ corrections to the entropy. Hence, to leading logarithmic accuracy,
\begin{align}
S_\ell
\simeq
\frac{\pi^2}{3}\langle \hat N_\ell^2\rangle_c
=
\frac{\pi^2}{6}G_{\ell\ell}.
\end{align}
Using the result above, this gives
\begin{align}
S_\ell
=
\frac{1}{6}\ln L+O(1),
\qquad
G_{\ell\ell}
=
\frac{6}{\pi^2}S_\ell+O(1).
\end{align}
Thus the Gaussian FCS exponent may be written equivalently in terms of either the relative-string fluctuation $G_{\ell\ell}$ or the entanglement entropy $S_\ell$.

Fig.~\ref{fig:small_lambda_structure} illustrates the two ingredients used in the mid-wall asymptotics. Fig.~\ref{fig:small_lambda_structure} (a) compares the exact finite-size FCS factor $\mathcal F(\lambda)$ with the Gaussian form $\exp[-G_{\ell\ell}\lambda^2/2]$. The agreement near $\lambda=0$ shows that the leading long-distance scale is controlled by the connected relative-string fluctuation $G_{\ell\ell}$, or equivalently by the midpoint entanglement entropy $S_\ell$. Fig.~\ref{fig:small_lambda_structure} (b) shows the hopping-insertion factor $B(\lambda)$. Its leading behavior is flat, $B(\lambda)=B(0)+O(\lambda^2)$, with $B(0)=2C_{\ell,\ell+1}$, confirming that the insertion mainly supplies the half-link phase shift while contributing only smooth prefactors to the asymptotic exponent.

\end{widetext}

\newpage
\begin{widetext}

\begin{center}
{\bf Supplemental Material}
\end{center}

In this Supplemental Material, we present: (i) the Feshbach--Schur method for deriving the effective potential; (ii) the full-wall virtual hopping operator and its joint-FCS representation;  (iii) Slater-determinant derivation of the FCS formulas; (iv) the DMRG methods used to compute the constrained ground-state energy.
\section{Feshbach--Schur method}
\label{sec:SM_feshbach}

In this section we derive the estimate used in the main text for the constrained ground-state energy shift. Let $P$ be a wall projector, either the full-wall projector or the mid-wall projector, and let $Q=1-P$. With respect to the decomposition $\mathcal H=P\mathcal H\oplus Q\mathcal H$, the unconstrained Hamiltonian has the block form
\begin{align}
H_0
=
\begin{pmatrix}
H_{PP} & H_{PQ}\\
H_{QP} & H_{QQ}
\end{pmatrix}
=
\begin{pmatrix}
PH_0P & PH_0Q\\
QH_0P & QH_0Q
\end{pmatrix}.
\end{align}
Let $H_0|\psi_0\rangle=E_0|\psi_0\rangle$, and decompose $|\psi_0\rangle=|p\rangle+|q\rangle$ with $|p\rangle=P|\psi_0\rangle$ and $|q\rangle=Q|\psi_0\rangle$. The Schrödinger equation gives
\begin{align}
(H_{PP}-E_0)|p\rangle+H_{PQ}|q\rangle&=0,\\
H_{QP}|p\rangle+(H_{QQ}-E_0)|q\rangle&=0.
\end{align}
Assuming that $H_{QQ}-E_0$ is invertible on the relevant $Q$ subspace, the second equation gives
\begin{align}
|q\rangle
=
-(H_{QQ}-E_0)^{-1}H_{QP}|p\rangle .
\end{align}
Substituting this into the first equation yields the Feshbach--Schur effective operator in the allowed sector,
\begin{align}
F_P(z)
=
H_{PP}
-
H_{PQ}(H_{QQ}-z)^{-1}H_{QP},
\end{align}
with $F_P(E_0)|p\rangle=E_0|p\rangle$.

It is useful to define the self-energy
\begin{align}
\Sigma_P(E_0)
=
H_{PQ}(H_{QQ}-E_0)^{-1}H_{QP}.
\end{align}
Then $H_{PP}=F_P(E_0)+\Sigma_P(E_0)$. Let $|\psi_p\rangle=|p\rangle/\sqrt{\langle p|p\rangle}$ be the normalized projected state. To leading order in the self-energy, we have
\begin{align}
E_+(r)
=
E_0+
\langle\psi_p|\Sigma_P(E_0)|\psi_p\rangle+\cdots .
\end{align}
Using the relation for $|q\rangle$, this matrix element can be rewritten as
\begin{align}
\langle\psi_p|\Sigma_P(E_0)|\psi_p\rangle
=
-\frac{\langle p|H_{PQ}|q\rangle}{\langle p|p\rangle}.
\end{align}
Since $\langle p|H_{PQ}|q\rangle=\langle\psi_0|H_{PQ}|\psi_0\rangle$ and $\langle p|p\rangle=1-\langle\psi_0|Q|\psi_0\rangle$, the leading estimate for the effective potential is
\begin{align}
\Delta E(r)
=
E_+(r)-E_0
\simeq
-
\frac{\langle\psi_0|H_{PQ}|\psi_0\rangle}
{1-\langle\psi_0|Q|\psi_0\rangle}.
\label{eq:SM_feshbach_estimate}
\end{align}
In the regime considered in the main text, the forbidden-sector weight $\langle\psi_0|Q|\psi_0\rangle$ is exponentially small, so the denominator gives only a smooth normalization factor. The leading separation dependence is therefore controlled by the wall-hopping matrix element $\langle\psi_0|H_{PQ}|\psi_0\rangle$.

\section{Full-wall problem and joint FCS}
\label{sec:SM_fullwall}

We now derive the full-wall virtual hopping operator and its joint-FCS representation. The full-wall projector is
\begin{align}
P
=
\prod_{j=1}^{L-1}\Theta(\hat u_j+r),
\qquad
Q=1-P,
\end{align}
where the physical configurations satisfy $u_j>-r$ for all $j$. In the fermion representation, $\hat u_j=\hat N_j^{(1)}-\hat N_j^{(2)}$ with $\hat N_j^{(a)}=\sum_{x=1}^{j}\hat n_{x,a}$.

A hop across bond $y$ changes only the relative-string coordinate $u_y$. The two terms that raise $u_y$ by one are
\begin{align}
T_y^{(+)}
=
c_{y,1}^{\dagger}c_{y+1,1}
+
c_{y+1,2}^{\dagger}c_{y,2}.
\end{align}
The first term moves a fermion on chain $1$ from the right side of bond $y$ to the left side, while the second term moves a fermion on chain $2$ from the left side to the right side. Both operations increase $\hat u_y$ by one and leave all other $\hat u_j$ unchanged.

Therefore a state in $Q$ can be hopped into $P$ across bond $y$ only if $u_y=-r$ before the hop, while all other coordinates already obey $u_j>-r$. This gives the exact full-wall wall-hopping operator
\begin{align}
H_{PQ}^{\rm full}
=
-t
\sum_{y=1}^{L-1}
T_y^{(+)}
\delta_{\hat u_y,-r}
\prod_{j\ne y}\Theta(\hat u_j+r).
\label{eq:SM_HPQ_full}
\end{align}
Consequently,
\begin{align}
\langle H_{PQ}^{\rm full}\rangle_0
=
-t
\sum_{y=1}^{L-1}
\left\langle
T_y^{(+)}
\delta_{\hat u_y,-r}
\prod_{j\ne y}\Theta(\hat u_j+r)
\right\rangle_0,
\label{eq:SM_HPQ_full_exp}
\end{align}
where $\langle\cdots\rangle_0$ denotes the expectation value in the unconstrained two-chain ground state.

For fixed $y$, define the boundary set
\begin{align}
\mathcal A_r^{(y)} = 
\left\{
\mathbf m=(m_1,\dots,m_{L-1})\in\mathbb Z^{L-1}:
m_y=-r,\quad m_j>-r\ {\rm for}\ j\ne y
\right\}.
\end{align}
Then
\begin{align}
\delta_{\hat u_y,-r}
\prod_{j\ne y}\Theta(\hat u_j+r)
=
\sum_{\mathbf m\in\mathcal A_r^{(y)}}
\delta_{\hat{\mathbf u},\mathbf m},
\qquad
\delta_{\hat{\mathbf u},\mathbf m}
=
\prod_{j=1}^{L-1}\delta_{\hat u_j,m_j}.
\end{align}
Using the Fourier representation
\begin{align}
\delta_{\hat{\mathbf u},\mathbf m}
=
\int_{[-\pi,\pi]^{L-1}}
\frac{d^{L-1}\lambda}{(2\pi)^{L-1}}
e^{-i\boldsymbol\lambda\cdot\mathbf m}
e^{i\boldsymbol\lambda\cdot\hat{\mathbf u}},
\end{align}
we obtain
\begin{align}
\langle H_{PQ}^{\rm full}\rangle_0
=
-t
\sum_{y=1}^{L-1}
\sum_{\mathbf m\in\mathcal A_r^{(y)}}
\int_{[-\pi,\pi]^{L-1}}
\frac{d^{L-1}\lambda}{(2\pi)^{L-1}}
e^{-i\boldsymbol\lambda\cdot\mathbf m}
\mathcal F(\boldsymbol\lambda)A_y(\boldsymbol\lambda).
\label{eq:SM_fullwall_FCS}
\end{align}
Here
\begin{align}
\mathcal F(\boldsymbol\lambda)
=
\left\langle
e^{i\sum_{j=1}^{L-1}\lambda_j\hat u_j}
\right\rangle_0
=
\chi_L(\boldsymbol\lambda)\chi_L(-\boldsymbol\lambda)
\end{align}
is the joint FCS generating function of the relative-string variables, and $A_y(\boldsymbol\lambda)$ is the hopping insertion associated with bond $y$.

The Gaussian expansion of the joint FCS near $\boldsymbol\lambda=0$ is
\begin{align}
\mathcal F(\boldsymbol\lambda)
=
\exp\left[
-\frac12
\sum_{j,k=1}^{L-1}
\lambda_jG_{jk}\lambda_k
+
O(\lambda^4)
\right],
\qquad
G_{jk}
=
\langle \hat u_j\hat u_k\rangle_c .
\label{eq:SM_joint_gaussian}
\end{align}
The insertion is smooth at the origin and satisfies $A_y(\mathbf 0)=2C_{y,y+1}$. Keeping the leading phase of the insertion gives $A_y(\boldsymbol\lambda)\simeq 2C_{y,y+1}e^{-i\lambda_y/2}$ in the sign convention of Eq.~\eqref{eq:SM_fullwall_FCS}. Thus the half-link shift acts only on the crossed coordinate $u_y$.

At leading exponential accuracy, the $y$th contribution is controlled by the constrained Gaussian cost for reaching the wall at coordinate $y$. If the single-active-coordinate saddle is feasible, the optimal profile is
\begin{align}
h_j^{*(y)}
=
\frac{G_{jy}}{G_{yy}},
\end{align}
and the action is
\begin{align}
I_y^*
=
\frac{1}{2G_{yy}}.
\end{align}
Thus the contribution from cut $y$ has the leading scale
\begin{align}
\exp\left[
-\frac{(r-1/2)^2}{2G_{yy}}
\right]
\end{align}
up to subleading factors. The dominant cut is the one with $G_{yy}=G_{\max}$, giving
\begin{align}
\Delta E_{\rm full}(r)
\sim
\exp\left[
-\frac{(r-1/2)^2}{2G_{\max}}
\right]
\times
{\rm subleading\ factors}.
\end{align}
For the half-filled chain, $G_{\max}=\pi^{-2}\ln L+O(1)$, so the full-wall problem has the same leading scale as the mid-wall reduction,
\begin{align}
\Delta E_{\rm full}(r)
\sim
\exp\left[
-\frac{\pi^2r^2}{2\ln L}
\right]
\times
{\rm subleading\ factors}.
\end{align}

\section{Slater-determinant derivation of the FCS formulas}
\label{sec:supp_fcs_derivation}

In this section we derive the determinant formulas for the one-string FCS generating function $\chi(\lambda)$ and the hopping-inserted generating function $\Gamma(\lambda)$. Let $|\mathrm{GS}\rangle$ be a free-fermion Slater determinant with $M$ occupied orbitals,
\begin{align}
|\mathrm{GS}\rangle
=
d_1^\dagger\cdots d_M^\dagger|0\rangle,
\qquad
d_a^\dagger=\sum_{x=1}^{L}\Phi_{xa}c_x^\dagger ,
\end{align}
where the $L\times M$ matrix $\Phi$ has orthonormal columns, $\Phi^\dagger\Phi=I_M$. The one-body correlation matrix is
\begin{align}
C_{xy}=\langle c_x^\dagger c_y\rangle .
\end{align}
For the real single-string ground state used in the main text, $C=\Phi\Phi^\dagger$ in the site basis.

We first recall the overlap formula for two Slater determinants. If
\begin{align}
|\Phi\rangle=d_1^\dagger\cdots d_M^\dagger|0\rangle,
\qquad
|\Psi\rangle=\tilde d_1^\dagger\cdots \tilde d_M^\dagger|0\rangle,
\end{align}
with occupied-orbital matrices $\Phi$ and $\Psi$, then
\begin{align}
\langle \Phi|\Psi\rangle
=
\det(\Phi^\dagger\Psi).
\label{eq:supp_slater_overlap}
\end{align}
This follows directly by commuting the annihilation operators in $\langle0|d_M\cdots d_1$ through the creation operators $\tilde d_1^\dagger\cdots\tilde d_M^\dagger|0\rangle$.

Now consider a diagonal counting field
\begin{align}
c^\dagger Xc=\sum_{x=1}^{L}X_x\hat n_x,
\qquad
D=e^X=\operatorname{diag}(e^{X_1},\dots,e^{X_L}).
\end{align}
The corresponding many-body operator transforms creation operators as
\begin{align}
e^{c^\dagger Xc}c_x^\dagger e^{-c^\dagger Xc}
=
e^{X_x}c_x^\dagger .
\end{align}
Therefore $e^{c^\dagger Xc}|\mathrm{GS}\rangle$ is again a Slater determinant, with occupied-orbital matrix $D\Phi$. Using Eq.~\eqref{eq:supp_slater_overlap},
\begin{align}
\chi[X]
:=
\langle \mathrm{GS}|e^{c^\dagger Xc}|\mathrm{GS}\rangle
=
\det(\Phi^\dagger D\Phi).
\label{eq:supp_chi_phi}
\end{align}
This can be rewritten in terms of the correlation matrix. Since $\Phi^\dagger D\Phi=I_M+\Phi^\dagger(D-I)\Phi$, the identity $\det(I+AB)=\det(I+BA)$ gives
\begin{align}
\chi[X]
=
\det\!\left[I_L+(D-I)\Phi\Phi^\dagger\right].
\end{align}
Using $C=\Phi\Phi^\dagger$ and the fact that $D$ is diagonal, this is equivalently
\begin{align}
\chi[X]
=
\det(I-C+CD).
\label{eq:supp_chi_general}
\end{align}
This is the determinant formula for the FCS generating function.

We next derive the inserted formula. Define the normalized transition density
\begin{align}
(C_X)_{xy}
=
\frac{
\langle\mathrm{GS}|c_x^\dagger c_y e^{c^\dagger Xc}|\mathrm{GS}\rangle
}{
\langle\mathrm{GS}|e^{c^\dagger Xc}|\mathrm{GS}\rangle
}.
\end{align}
Because the ket $e^{c^\dagger Xc}|\mathrm{GS}\rangle$ has occupied-orbital matrix $D\Phi$, the transition density between the bra Slater determinant $\Phi$ and the ket Slater determinant $D\Phi$ is
\begin{align}
C_X
=
\Phi(\Phi^\dagger D\Phi)^{-1}\Phi^\dagger D .
\label{eq:supp_transition_density}
\end{align}
This expression can be obtained by expanding $c_y$ on the occupied orbitals of the ket Slater determinant and using the inverse overlap matrix $(\Phi^\dagger D\Phi)^{-1}$ to contract with the bra orbitals. Equivalently, it follows from the cofactor expansion of the Slater overlap after replacing one occupied orbital by the site basis vector selected by $c_x^\dagger c_y$.

We now express Eq.~\eqref{eq:supp_transition_density} in terms of $C$. Let
\begin{align}
M_X=I-C+CD.
\end{align}
Using $C=\Phi\Phi^\dagger$ and $S=\Phi^\dagger D\Phi$, one checks that
\begin{align}
M_X\Phi S^{-1}\Phi^\dagger D
=
CD .
\end{align}
Therefore, whenever $M_X$ is invertible,
\begin{align}
C_X
=
M_X^{-1}CD
=
(I-C+CD)^{-1}CD .
\label{eq:supp_CX_general}
\end{align}
Combining Eqs.~\eqref{eq:supp_chi_general} and \eqref{eq:supp_CX_general}, we obtain the inserted determinant formula
\begin{align}
\langle\mathrm{GS}|c_x^\dagger c_y e^{c^\dagger Xc}|\mathrm{GS}\rangle
=
\det(I-C+CD)
\left[
(I-C+CD)^{-1}CD
\right]_{xy}.
\label{eq:supp_inserted_general}
\end{align}

For the mid-wall problem, the counting field is $X=i\lambda P_\ell$, where $P_\ell$ projects onto sites $1,\dots,\ell$. Thus,
\begin{align}
D_\ell(\lambda)
=
I+(e^{i\lambda}-1)P_\ell .
\end{align}
The one-string FCS generating function is
\begin{align}
\chi(\lambda)
=
\langle e^{i\lambda\hat N_\ell}\rangle
=
\det[I-C+CD_\ell(\lambda)].
\label{eq:supp_chi_mid_final}
\end{align}
The hopping-inserted generating function is
\begin{align}
\Gamma(\lambda)
=
\langle c_\ell^\dagger c_{\ell+1}e^{i\lambda\hat N_\ell}\rangle
=
\chi(\lambda)m(\lambda),
\end{align}
with
\begin{align}
m(\lambda)
=
\left[
(I-C+CD_\ell(\lambda))^{-1}CD_\ell(\lambda)
\right]_{\ell,\ell+1}.
\label{eq:supp_m_mid_final}
\end{align}
These are the formulas used in the main text.

For the full-wall problem, the one-string joint counting field is
\begin{align}
\exp\left(i\sum_{j=1}^{L-1}\lambda_j\hat N_j\right),
\qquad
\hat N_j=\sum_{x=1}^{j}\hat n_x .
\end{align}
The exponent is site diagonal:
\begin{align}
\sum_{j=1}^{L-1}\lambda_j\hat N_j
=
\sum_{x=1}^{L}\alpha_x\hat n_x,
\qquad
\alpha_x=\sum_{j=x}^{L-1}\lambda_j,
\qquad
\alpha_L=0.
\end{align}
Thus
\begin{align}
D(\boldsymbol\lambda)
=
\operatorname{diag}(e^{i\alpha_1},\dots,e^{i\alpha_L}),
\end{align}
and Eq.~\eqref{eq:supp_chi_general} gives
\begin{align}
\chi_L(\boldsymbol\lambda)
=
\det[I-C+CD(\boldsymbol\lambda)].
\label{eq:supp_chi_full_final}
\end{align}
Defining
\begin{align}
R(\boldsymbol\lambda)
=
[I-C+CD(\boldsymbol\lambda)]^{-1}CD(\boldsymbol\lambda),
\end{align}
the hopping insertion for a hop across bond $y$ is
\begin{align}
A_y(\boldsymbol\lambda)
=
R(\boldsymbol\lambda)_{y,y+1}
+
R(-\boldsymbol\lambda)_{y+1,y}.
\label{eq:supp_Ay_full_final}
\end{align}
The first term corresponds to $c_y^\dagger c_{y+1}$ on chain $1$, and the second term corresponds to $c_{y+1}^\dagger c_y$ on chain $2$ with the opposite counting field.

Finally, we record the small-counting-field expansion of the midpoint insertion. With $D_\ell(\lambda)=I+\alpha P_\ell$ and $\alpha=e^{i\lambda}-1$,
\begin{align}
I-C+CD_\ell(\lambda)=I+\alpha CP_\ell,
\end{align}
so
\begin{align}
(I-C+CD_\ell)^{-1}CD_\ell
=
C+\alpha(CP_\ell-CP_\ell C)+O(\alpha^2).
\end{align}
Taking the $(\ell,\ell+1)$ matrix element and using $(CP_\ell)_{\ell,\ell+1}=0$, one obtains
\begin{align}
m(\lambda)
=
C_{\ell,\ell+1}
-
i\lambda(CP_\ell C)_{\ell,\ell+1}
+
O(\lambda^2).
\end{align}
For the real single-string ground state, $m(-\lambda)=m(\lambda)^*$, so
\begin{align}
A(\lambda)=m(-\lambda)+e^{i\lambda}m(\lambda)=e^{i\lambda/2}B(\lambda),
\qquad
B(\lambda)=2C_{\ell,\ell+1}+O(\lambda^2).
\end{align}
This is the origin of the half-link shift in the midpoint FCS integral.

\section{DMRG with an explicit hard-core projector}
\label{sec:supp_dmrg}

For larger systems, we use DMRG for the projected Hamiltonian
\begin{align}
H_+=PH_0P .
\end{align}
The two XY chains are represented as a two-leg ladder. At each position $j$, the local Hilbert space consists of the four rung states
\begin{align}
|UU\rangle,\quad |UD\rangle,\quad |DU\rangle,\quad |DD\rangle,
\end{align}
where the two entries denote the local spin states on chains $1$ and $2$. The unconstrained Hamiltonian $H_0$ is represented as an MPO containing nearest-neighbor XY exchange terms on each leg.

\begin{figure}[t]
  \centering
  \includegraphics[width=0.38\textwidth]{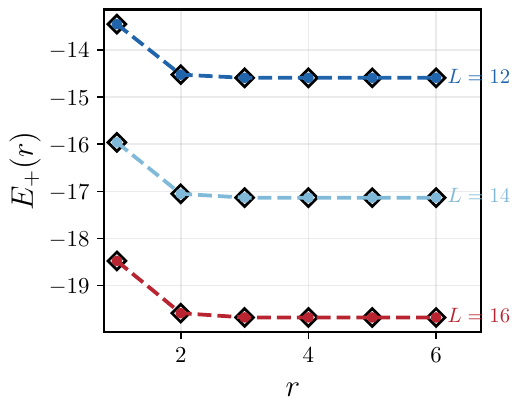}
  \caption{
  Comparison of constrained ground state energies $E_+(r)$ from ED and DMRG for $L=12,14,16$. Black hollow diamonds show ED values obtained from $E_+(r)=E_0+\Delta E(r)$, while colored dashed lines with solid dots show DMRG energies computed directly from the projected Hamiltonian $H_+=PH_0P$.
  }
  \label{fig:supp_ed_dmrg_eplus}
\end{figure}

The hard-core constraint is implemented by an explicit diagonal MPO for the projector $P$. The virtual index of this MPO stores the running value of the relative separation. Equivalently, it records the accumulated relative string variable as one sweeps from left to right. At each rung, the virtual state is updated according to the local spin configuration, and transitions that would violate $u_j>-r$ are removed. The endpoint condition fixes the final accumulated displacement, corresponding to the fixed-endpoint sector. In this way the MPO implements the full product projector
\begin{align}
P=\prod_{j=1}^{L-1}\Theta(\hat u_j+r)
\end{align}
without introducing a penalty term.

The DMRG calculation is then performed on $H_+=PH_0P$. During the sweeps we monitor the projector expectation value
\begin{align}
\langle P\rangle_\psi=\langle\psi|P|\psi\rangle .
\end{align}
For a perfectly constrained state, $\langle P\rangle_\psi=1$. This quantity therefore provides a direct diagnostic of leakage from the allowed Hilbert space due to MPO compression and MPS truncation. The calculations reported in the main text are converged with respect to bond dimension and truncation cutoff; the residual constraint violation is negligible on the scale of the plotted energy differences.

As a small-system check of the projected-MPO implementation, Fig.~\ref{fig:supp_ed_dmrg_eplus} compares the constrained ground state energies $E_+(r)$ from DMRG with ED for $L=12,14,16$ and $r=1,\ldots,6$. The agreement of the data verifies that the explicit hard-core projector reproduces the constrained spectrum in the regime where ED is available.

\end{widetext}

\end{document}